\documentclass[journal]{IEEEtran}

\usepackage{graphicx}
\usepackage{subfigure}  
\usepackage{diagbox}  
\usepackage{epstopdf} 
\usepackage{multirow}
\usepackage{cite}
\usepackage{stfloats}
\usepackage{amsmath}
\usepackage[square, comma, sort&compress, numbers]{natbib}
\usepackage{setspace} 
\newcounter{mytempeqncnt}  

\begin{document}

\title{Non-Orthogonal Random Access (NORA) for 5G Networks}

\author{Yanan~Liang,~\IEEEmembership{Student~Member,~IEEE,}
        Xu~Li,~\IEEEmembership{Member,~IEEE,}
        Jiayi~Zhang,~\IEEEmembership{Member,~IEEE}
        and~Zhiguo~Ding,~\IEEEmembership{Senior Member,~IEEE}
\thanks{This work was supported in part by the National High-tech Research and Development Program of China (863 Program) under Grant 2015AA01A709, the National Natural Science Foundation
of China (Grant Nos. 61601020 and 61371068), and the Fundamental Research
Funds for the Central Universities (Grant Nos. 2017JBM319, 2016RC013, 2014JBZ002 and 2016JBZ003). The work of Z. Ding was supported by the UK EPSRC under grant number EP/L025272/1 and by H2020-MSCA-RISE-2015 under grant number 690750. Parts of this paper was presented at the IEEE Wireless Communications and Networking Conference (WCNC), 19-22 March, 2017, San Francisco, CA. (Corresponding author: Jiayi Zhang.)}
\thanks{Y. Liang, X. Li and J. Zhang are with the School of Electronic and Information Engineering, Beijing Jiaotong University, Beijing 100044, P. R. China (e-mail: {14111032, xli, jiayizhang}@bjtu.edu.cn)}
\thanks{Z. Ding is with the School of Computing and Communications, Lancaster University, LA1 4YW, U.K. (e-mail: z.ding@newcastle.ac.uk).}
}

%



\maketitle

\begin{abstract}
The massive amounts of machine-type user equipments (UEs) will be supported in the future fifth generation (5G) networks. However, the potential large random access (RA) delay calls for a new RA scheme and for a detailed assessment of its performance. Motivated by the key idea of non-orthogonal multiple access, the non-orthogonal random access (NORA) scheme based on successive interference cancellation (SIC) is proposed in this paper to alleviate the access congestion problem. Specifically, NORA utilizes the difference of time of arrival to identify multiple UEs with the identical preamble, and enables power domain multiplexing of collided UEs in the following access process, while the base station performs SIC based on the channel conditions obtained through preamble detection. Our analysis show that the performance of NORA is superior to the conventional orthogonal random access (ORA) scheme in terms of the preamble collision probability, access success probability and throughput of random access. Simulation results verify our analysis and further show that our NORA scheme can improve the number of the supported UEs by more than 30\%. Moreover, the number of preamble transmissions and the access delay for successfully accessed UEs are also reduced significantly by using the proposed random access scheme.  
\end{abstract}

\begin{IEEEkeywords}
Random access, collision probability, throughput, access delay.
\end{IEEEkeywords}

%
\IEEEpeerreviewmaketitle

\section{Introduction}
%
%
%
%
\IEEEPARstart{T}{he} fifth generation (5G) networks will support tens of thousands user equipments (UEs) per cell in the near future \cite{17NORA,16andrews2014will}. Each UE performs a random access (RA) procedure for initial uplink access to connect and synchronize with its base station \cite{12TS36.300}. When the number of UEs is tremendous, the RA procedure is inefficient due to the frequent transmission collisions, which lead to network congestion, unexpected delay, high power consumption, and radio resource wastage. Hence, the RA procedure becomes the bottleneck of 5G networks' performance \cite{9wei2015modeling}. 

In current LTE systems, the RA procedure consists of a four-message handshake between the UE and the eNodeB (which is referred to as orthogonal random access (ORA) scheme in the following sections). The four messages include Preamble, Random Access Response (RAR), Initial Layer 3 message (Msg3) and Contention Resolution (CR) \cite{1TS36.321}. A periodic sequence of time-frequency resources called random access slots (RA slots) are reserved in the Physical Random Access Channel (PRACH) for preamble transmission \cite{2TS36.211}. Whenever a UE triggers the RA procedure, it transmits a preamble randomly chosen from the available orthogonal pseudo-random preambles periodically broadcast by the eNodeB in the next available RA slot. There are up to 64 available preambles within each cell \cite{7LTE}. So if more than 64 UEs make RA attempts in one specific RA slot, the collision is inevitable. The UEs whose preambles have been successfully recognized by the eNodeB will receive RAR and transmit Msg3 on the Physical Uplink Shared Channel (PUSCH). In 3GPP RA operation, collision happens if more than one UE select the same preamble. The UEs that experience collision will not be scheduled for Msg3 transmission, and will return to preamble transmission \cite{11laya2014random}. As the number of UEs grows, the collision becomes more and more frequent and finally leads to congestion. The UEs end up transmitting preambles repeatedly until the maximum allowed number of preamble transmissions is reached. Then the UEs declare access failure and exit the RA procedure. Even if the UEs manage to successfully complete the RA procedure within the maximum allowed number of preamble transmissions, the access delay may still be intolerable. The congestion will block most of the RA attempts from UEs even if the network has lots of unused radio resource, and leads to under-utilized networks.

\subsection{Related Work}
Several solutions have been proposed to handle the RA congestion problem in pioneering works, such as access class barring (ACB) \cite{13wang2015optimal,wiriaatmadja2015hybrid,duand,morvari2016two,seo2011design}, extended access barring (EAB) \cite{lin2014prada}, dynamic allocation \cite{14pang2014network}, specific backoff scheme \cite{15chen2015delayed}, and pull-based scheme \cite{6TR37.868}. 

By introducing a separate access class, ACB allows the eNodeB to control the access of UEs separately. Two vital parameters in the ACB method are the barring factor which represents the probability of barring and the backoff factor which indicates the backoff time before retrying random process if the UE fails the ACB check. Many scholars have worked on the dynamic adjustment of the barring factor. In \cite{wiriaatmadja2015hybrid}, a joint resource allocation and access barring scheme is proposed to achieve uplink scheduling and random access network (RAN) overload control, in which the access barring parameter is adaptively changed based on the amount of available RBs and the traffic load. In \cite{duand}, two dynamic ACB algorithms for fixed and dynamic preamble allocation schemes are proposed to determine the barring factors without priori knowledge of the number of MTC devices. \cite{13wang2015optimal} formulates an optimization problem to determine the optimal barring parameter which maximizes the expected number of MTC devices successfully served in each RA slot. \cite{morvari2016two} proposes a two-stage ACB scheme to increase access success probability. In the first stage, the UEs use the barring factor broadcast by the eNodeB. The UEs which pass the ACB check are viewed as primary UEs and allowed to select non-special preambles randomly, while the UEs which fail are treated as secondary UEs and select the special preambles. In the second stage, each secondary UE calculates its barring probability independently based on the expected number of secondary UEs.
In terms of the backoff factor, \cite{seo2011design} compares the performance of uniform backoff (UB) and binary exponential backoff (BEB) algorithms and proposes a new algorithm to adaptively adjust the backoff window size under unsaturated traffic conditions. 

EAB extends the granularity of the access class to distinguish multiple classes, which has been introduced in 3GPP standard to throttle the access of Machine Type Communication (MTC) devices \cite{6TR37.868}. A prioritized random access with dynamic access barring (PRADA) framework is proposed in \cite{lin2014prada}, which optimizes the EAB parameters such as activation time, barring opportunity, and backoff time. PRADA includes two components: pre-allocation of different amount of RA slots for different classes and dynamic access barring (DAB). The average number of successful preambles is observed in each two neighboring RA slots to estimate the RA load. If the RA load is heavy, EAB is triggered and the RA attempts of the UEs with their first preamble transmissions are deferred for a long time. However, there are concerns that EAB may be frequently activated and deactivated since the number of successful preambles drastically varies for bursty RA arrivals. It will result in performance deterioration, for instance, decrease in RA success probability. 

To guarantee throughput, \cite{14pang2014network} proposes a game-theoretic framework to dynamically allocate additional RA resources to MTC devices. \cite{15chen2015delayed} elaborates a MTC specific backoff scheme which introduces a dynamic backoff indicator assignment algorithm to reduce RA collision probability. The pull-based scheme \cite{6TR37.868} allows the eNodeB to control network load by dominating the paging operation, where MTC devices will trigger RA process upon receiving a paging message.

The main idea behind aforementioned schemes is to disperse the transmission of the access request to control overload and increase the access probability within a relatively short time. In spite that these approaches can reduce access collision to a certain degree, the retransmission of numerous UEs can again aggravate collision and further increase the access delay. 

Moreover, several schemes have been developed to mitigate collision and reduce access delay by increasing available RA resources. For example, Thomsen et al. \cite{5Thomsen14code} proposes a code-expanded RA scheme which adopts the concept of access codeword to increase the amount of available contention resources in the RA process. Moreover, a preamble reuse scheme is proposed in \cite{4ko14spatial}, which spatially partitions the cell coverage into multiple regions and reduces the cyclic shift size to generate more preambles. However, the extent of the RA resource increase is limited and severe collision and retransmission are still inevitable. 

The RAN overload problem can also be tackled by efficiently utilizing the radio resources for the RA process. In \cite{3ko2012novel}, the authors propose a novel random access scheme based on fixed timing alignment information to reduce collision probability given a large number of fixed-location machine-to-machine (M2M) devices. In the scheme proposed in \cite{niyato2014performance}, M2M UEs form coalitions and perform relay transmission with an objective to reduce network congestion.
To increase the preamble detection probability, the authors of \cite{10kim2015enhanced} propose an enhanced RA scheme, in which the eNodeB adopts the transmission time difference to detect the UEs which utilizes the same preamble. Furthermore, the eNodeB creates multiple RARs in response to detected UEs which select the same preamble. However, this scheme fails to consider the limitation on PUSCH resources. With the increase of successfully transmitted preambles, the limited PUSCH resources may be another bottleneck for RA process.

In summary, most of existing solutions to improve LTE RA performance mainly focus on controlling traffic or increasing available RA resources. However, the available RA resources are limited. Moreover, the access delay cannot be guaranteed by controlling traffic.

Successive interference cancellation (SIC) enables throughput efficiency enhancements by utilizing collided packets for decoding instead of discarding them \cite{Yu2008High}, which makes it an obvious candidate for alleviating RA congestion problem. 
\cite{Salvo2015Power} evaluates the random access throughput performance of asynchronous code division multiple access (CDMA) systems with interference cancellation receivers. Power randomization is explored to aid iterative receiver processing. \cite{Applebaum2012Asynchronous} proposes a code-division random access (CDRA) scheme, which adopts specific sets of codewords to spread the uplink resources in a non-orthogonal manner among users. The codewords assignment scheme allows a random subset of users communicating single bits to the base station (BS). It is worth noting that CDRA uses a convex optimization-based multiple user detection algorithm to avoid obtaining the delays and channel state information (CSI) of the users at the BS.  
Moreover, \cite{Yu2008High,Wang2008A,Wang2008Design} employ SIC to resolve collisions in the tree (also known as splitting) algorithm to improve random access throughput. These schemes rely on the property of tree algorithm where all packets are retained one-by-one in line with the underlying tree structure. However, this property is not available in LTE RA assumption. 
There also has been extensive studies on random access networks with SIC-enabled multi-packet reception capabilities. Nevertheless, most of recent works focus on the theoretical analysis of capture probability \cite{Zanella2012Theoretical} and power allocation scheme design \cite{Lin2015A,Xu2013Decentralized}. Few studies have considered the backoff and retransmission process in LTE RA procedure. To the best of our knowledge, systematic performance analysis of the SIC-enabled RA process under LTE scenario has not been reported yet. 
Recently, non-orthogonal multiple access (NOMA) has received many interests \cite{ding2014performance,ding2015cooperative,zhang2016uplink,liu2016cooperative}. Combined with SIC, NOMA allows simultaneously transmissions of multiple UEs with different powers \cite{ding2015cooperative}. Nonetheless, recent studies on NOMA mainly focus on performance analysis of data transmission process. In practical networks such as LTE-A and future 5G networks, the introduction and realization of NOMA and SIC in the random access process could be very challenging. As far as we're concerned, corresponding protocol and mechanism design which is compatible with existing LTE standard has not been reported yet.

\subsection{Motivation and Contribution}
Massive-connections, high-reliability and low-latency are typical technical scenarios for Internet of Things (IoT) in the 5G network. Specifically, smartphones, tablets and M2M communications such as environmental monitoring and smart meter reading applications have generated an explosive growth in the number of UEs. The surging RA attempts of enormous UEs call for more efficient and robust RA mechanisms. 
Motivated by the idea of NOMA and SIC, we propose a SIC-based non-orthogonal random access (NORA) mechanism which is easily applicable to existing LTE standard and future 5G standards. In contrast to ORA scheme, NORA facilitates the simultaneous transmission of Msg3 of collided UEs instead of conducting retransmission of preambles, which avoids further exacerbating collision without increasing demands on PUSCH resources. The information of UE locations and channel conditions is utilized to realize power domain multiplexing on the UE side and SIC on the base station side. Hereafter, we investigate the performance of NORA in terms of throughput, preamble collision probability, access success probability, access delay, and the number of preamble transmissions. Simulation results show that our NORA scheme can improve the number of the supported UEs by more than 30\%. The number of preamble transmissions and the access delay for successfully accessed UEs are also reduced remarkably.  

The contributions of this paper are summarized as follows:
\begin{itemize}
\item We propose a novel NORA mechanism which utilizes the spatial distribution characteristic of UEs. NORA integrates the arrival time-based multi-preamble detection and distance-based RAR reception schemes to effectively improve the preamble transmission success probability. The corresponding RAR message format is tailored for practical realizations. 
\item We derive an analytical model to investigate the transient behavior of the NORA process with non-stationary arrivals. Realistic assumptions, such as UEs' positions and channel conditions, are considered to show practical results. 
\item We provide a comprehensive analysis of the NORA performance, including throughput of preamble transmission and random access, collision/access success probability, the cumulative distribution function (CDF) of the number of preamble transmissions and access delay for the successfully accessed UEs, average number of preamble transmissions and average access delay. 
\item With the proposed NORA scheme, the throughput of random access process is increased by more than 30\%. Moreover, the average number of preamble transmissions and average access delay are reduced remarkably, and are only half as much as those of the ORA scheme in the best case. Comparisons with state-of-the art EAB scheme further validate the superiority of the NORA scheme. 
\end{itemize}

The rest of the paper is organized as follows. A detailed description of the NORA mechanism is presented in Section II. Section III depicts the analytical model of the proposed NORA procedure. The performance metrics and simulation analysis are given in Section IV, while Section V concludes the paper.

\section{Non-orthogonal random access mechanism}
In this section, we give a detailed description of the NORA scheme, which consists of PRACH preamble transmission, random access response, initial layer 3 message transmission and contention resolution (as illustrated in Fig. 1).

\begin{figure*}[!t]
\centering
\includegraphics[width=0.8\linewidth]{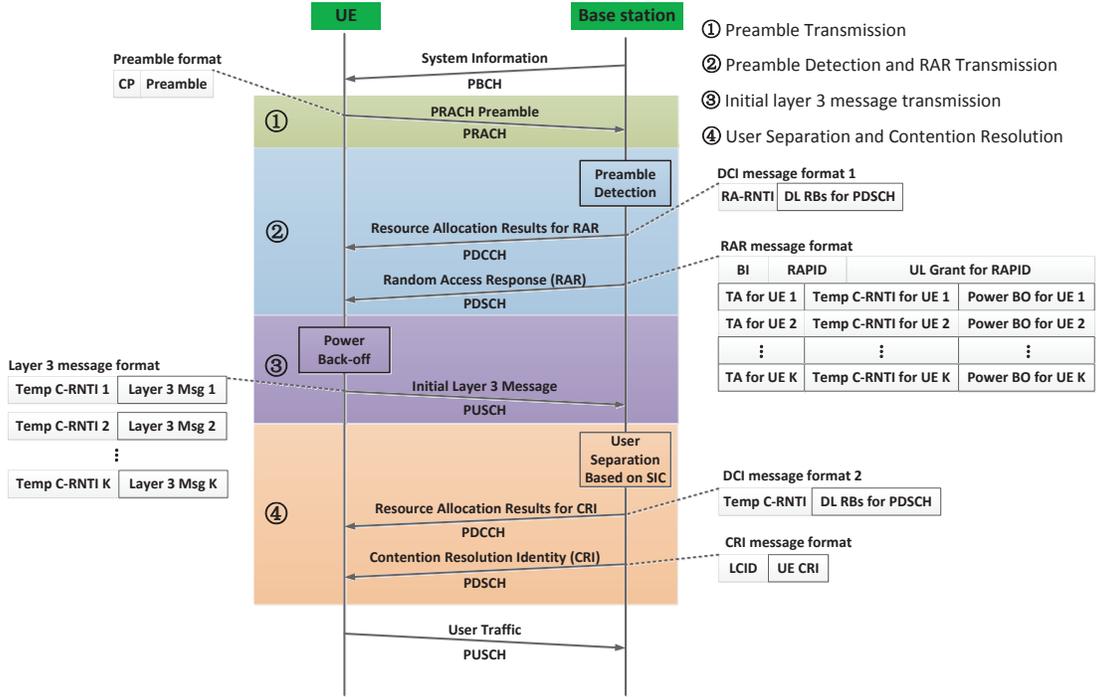}
\caption{Non-orthogonal Random Access Process.}
\label{Fig. 1}
\end{figure*}

\subsection{Preamble Transmission}\textbf{}
Each UE first receives the system information broadcast on Physical Broadcast Channel (PBCH) and acquires necessary configuration information to complete the RA process \cite{1TS36.331}. The information includes PRACH configuration information such as PRACH Configuration Index, PRACH Frequency Offset, Root Sequence Index, etc. and RACH configuration information such as Number of RA Preambles, Maximum Number of Preamble Transmission, RA Response Window Size, Power Back-off Offset, MAC Contention Resolution Timer, etc. When a UE starts to perform random access, it randomly selects a preamble sequence from the available preambles broadcast by the base station and transmits it in the next available RA slot. Preamble sequences are identified by their Random Access Preamble Identity (RAPID). There is also a one-to-one mapping between Random Access Radio Network Temporary Identifier (RA-RNTI) and the time/frequency resources used by the PRACH preamble. 

\subsection{Preamble Detection and RAR transmission}
\subsubsection{Arrival time based multi-preamble detection}
The base station first extracts the relevant PRACH signals within specific time/frequency resources through time-domain sampling and frequency-tone extraction. Then the base station computes the PRACH preamble power delay profile (PDP) through frequency-domain periodic correlation. Since different PRACH preambles are generated from cyclic shifts of a common root sequence, the periodic correlation operation provides in one shot the concatenated PDPs of all preambles derived from the same root sequence, as shown in Fig. 2.  

Each cyclic shift defines a Zero Correlation Zone (ZCZ), i.e. detection zone for corresponding preamble. The preamble detection process consists of searching the PDP peaks above a detection threshold within each ZCZ. The length of each ZCZ is determined by the cell size. When the cell size is more than twice the distance corresponding to the maximum delay spread, the base station may be able to differentiate the PRACH transmissions of two UEs which select the same preamble since they appear distinctly apart in the PDP (see Scenario 2 in Fig. 3), i.e. detect collision \cite{7LTE}. The Timing Advance (TA) value is calculated based on the time of arrival $\tau$.

\begin{figure}[!t]
\centering
\includegraphics[width=0.9\linewidth]{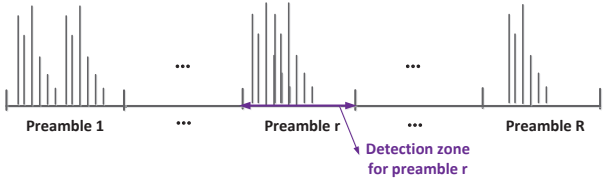}
\caption{Power delay profile of preambles within specific time/frequency resources.}
\label{Fig. 2}
\end{figure}

\begin{figure}[!t]
\centering
\includegraphics[width=0.9\linewidth]{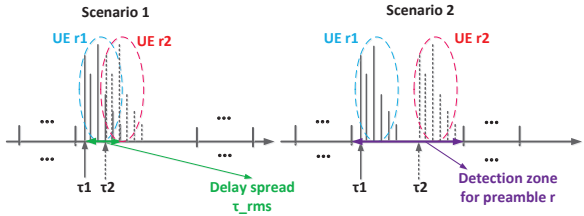}
\caption{Preamble detection scenarios when two UEs select the same preamble.}
\label{Fig. 3}
\end{figure}

\subsubsection{Transmission of DL RB for RAR}
Each UE can be differentiated by its RN-RNTI and RAPID. After transmitting the PRACH preamble, the UE searches for a Physical Downlink Control Channel (PDCCH) with its RA-RNTI within the response window, which starts from the third subframe after the PRACH transmission and its length is defined by RA Response Window Size. The PDCCH contains the downlink resource blocks (DL RB) and directs the UE to the associated RAR within the Physical Downlink Shared Channel (PDSCH). If the UE fails to find a PDCCH with its RA-RNTI, it will return to PRACH preamble transmission.

\subsubsection{Distance based RAR reception}
In ORA scheme, if the base station detects a preamble collision, it would not send any response regarding this specific RAPID, and the colliding UEs would randomly reselect their preambles and retransmit. However, in NORA, the eNodB would consider the UEs with detected collisions as a NORA group and respond to the group of UEs with a RAR message with the format shown in Fig. 1. 
The Backoff Indicator (BI) field is used to signal the backoff timer while The RAPID field addresses the UEs by the index of their preamble sequence. If the UE identifies an RAPID field with an appropriate preamble identity, it reads the corresponding instance of the UL Grant and identifies the allocated uplink RBs for transmission on the PUSCH. Each UE estimates its distance from the base station based on the received signal strength of Cell-specific Reference Signal (CRS) to obtain the approximate scope of TA \cite{7LTE}. Based on the approximation of TA, each UE searches the closest TA value in the RAR payload and acquires the corresponding Temporary Cell Radio Network Temporary Identifier (Temp C-RNTI) and Power Back-Off (Power BO) indication. 

\subsection{Initial Layer 3 Message Transmission}

\subsubsection{UE power back-off}
Conventional power control schemes attempt to maintain a constant received power at the base station from different UEs. The uplink transmit power (in dBm) for PUSCH transmission is given by \cite{6TR36.213} 
\begin{equation}
{P_{U}} = \min \left\{ {{P_{Umax}},{P_{O\_U}} + 10{{\log }_{10}}\left( {{M_{U}}} \right) + \alpha PL} \right\},
\end{equation}
where ${P_{Umax}}$ is the maximum transmit power and ${P_{O\_U}}$ represents the received power per RB when assuming a path loss of 0dB. ${M_{U}}$ denotes the number of available resource blocks in UL Grant while $PL$ denotes the downlink path loss estimate. $\alpha$  represents the reduced rate of transmit power increase due to fractional power control \cite{6TR36.213}.

In NORA, the base station performs user separation based on SIC, which requires diverse arrived power of UEs. To obtain this, the power back-off scheme \cite{zhang2016uplink} is introduced. The transmit power of the $i$-th UE in a NORA group is expressed by
\begin{equation}
{P_{U,i}} =  \min \left\{ {P_{Umax}}, {{P_{O\_U}} \!-\! \left( {i \!-\! 1} \right)\delta \!+\! 10{{\log }_{10}}\left( {{M_{U,i}}} \right) \!+\! \alpha P{L_i}} \right\},
\label{transmit_power}
\end{equation}
where $\delta$ is the Power Back-off Offset broadcast on PBCH. 

Eq. \eqref{transmit_power} implies that the received power of UEs in a NORA group gradually degrades with a step of $\delta$, which is beneficial to cancel the co-channel interference successively. Since the case that ${P_{U,i}}$ is the same as ${P_{Umax}}$ only occurs in extreme cases, which are not the focus of this paper, the transmit power of the $i$-th UE in a NORA group is assumed to be 
\begin{equation}
{P_{U,i}} = {P_{O\_U}} - \left( {i - 1} \right)\delta  + 10{\log _{10}}\left( {{M_{U,i}}} \right) + \alpha P{L_i}.
\label{transmit_power_i}
\end{equation}

The power back-off order of UEs in a NORA group is decided according to the TA value. The UE with a larger TA will be assigned a larger order $i$, which indicates that the corresponding received power ${P_{U,i}}$ is smaller. 

\subsubsection{Initial layer 3 message transmission with power domain multiplexing}
After power back-off and timing alignment with base station, the UEs in a NORA group will transmit their respective initial layer 3 messages within the same RBs of PUSCH. The initial layer 3 message contains information of UE's Temp C-RNTI. 

\subsection{User Separation and Contention Resolution}
\subsubsection{User separation based on SIC}
The decoding order of UEs in a NORA group is consistent with the power back-off order, i.e. the UE with the strongest received power will be decoded first, which is reasonable due to the SIC assumption. It's easily seen that the Power BO indication in RAR already implies the decoding order of UEs in a NORA group, thus unlike \cite{zhang2016uplink}, the base station doesn't need an individual control channel to inform the UEs of the assigned order. After decoding the UE's message, the base station identifies the temp C-RNTI and receives the initial layer 3 message.

\subsubsection{CRI transmission}
The final stage of the RA procedure is contention resolution, which is based upon the base station responding with a Contention Resolution Identity (CRI) message. Similar to the second stage， the base station first transmits a Downlink Control Information (DCI) message to specify the DL RBs for PDSCH transmission. The PDSCH includes the Logical Channel Identity (LCID) Medium Access Control (MAC) subheader and CRI MAC control element, as shown in Fig. 1. The LCID field is used to identify the subsequent CRI MAC control element while the CRI MAC control element is utilized to reflect the initial layer 3 message sent by the UE. The CRI contains configuration information regarding subsequent data transmission.

The UE starts a CR timer (length determined by MAC Contention Resolution Timer broadcast on PBCH) after transmitting the initial layer 3 message. If the UE does not receive a response from the base station before the timer expires, it will return to the procedure of transmitting PRACH preambles.

\section{Analytical model for NORA}


The parameters of the NORA process considered in this paper are summarized in Table \ref{Summary of Parameters}. $U$ depicts the total number of UEs in the cell. Assume that all UEs attempt to access the network over a period of time $T_{AP}$  with the arrival distribution $p(t)$ \cite{6TR37.868}.

\begin{table}[!tp]  
\caption{Summary of Parameters}
\newcommand{\tabincell}[2]{\begin{tabular}{@{}#1@{}}#2\end{tabular}}
\centering          
\begin{tabular}{cp{4.5cm}p{1.9cm}}         
\hline
Notation & Meaning & Value\\ \hline        
$U$ & Total number of UEs & \tabincell{l}{5000,10000,\\ \textbf{20000},30000,\\ \textbf{40000},50000}\\ \hline       
$p(t)$ & Arrival distribution & Uniform/Beta distribution \\ \hline        
${T_{AP}}$ & Arrival period & 10s\\ \hline 
$T_{RAP}$ & RA slot period & 5ms\\ \hline

$d_c$ & Cell radius & 500m\\ \hline 
$t_{rms}$ & RMS of the delay spread & 0.3us\\ \hline
$R$ & Number of available preambles in each RA slot & 54\\ \hline
$L$ & The maximum number of preamble transmissions for each UE & 10\\ \hline 
$p_l$ & Preamble detection probability of the $l$-th preamble transmission & $1-e^{-l}$\\ \hline

$\hat{R_j}$ & Target data rate of message transmission for the $j$-th UE in a NORA group & 1.6\\ \hline
$\hat{R_0}$ & Target data rate of message transmission for the UE with no collision & 1.6\\ \hline
$\delta$ & Power back-off offset for PUSCH transmission & 3dB\\ \hline
$\frac{{{P_M}}}{{{\sigma ^2}}}$ & Target arrived signal noise ratio (SNR) & 10dB\\ \hline
$\theta$ & Standard deviation of Rayleigh distribution & 1\\ \hline

$T_{PRACH}$ & Transmission time for PRACH preamble & 1,\textbf{2},3ms\\ \hline 
$T_{PD}$ & Processing time for preamble detection at the base station & 2ms\\ \hline 
$T_{RAR}$ & Transmission time for RAR message & 1ms\\ \hline 
$T_{R-3}$ & Processing time between receiving RAR and sending Layer 3 message & 3ms\\ \hline 
$T_{Msg3}$ & Transmission time for Layer 3 message & 3ms\\ \hline
$T_{CR}$ & Transmission time for CR message & 1ms\\ \hline
$W_{RAR}$ & Length of the RA response window & \tabincell{l}{2,3,4,5,\textbf{6},7,\\8,10ms}\\ \hline 
$W_{CR}$ & Length of the CR timer & \tabincell{l}{8,\textbf{16},24,32,40,\\48,56,64ms}\\ \hline
$W_{BO}$ & Length of the backoff window & 20ms\\ \hline
$\eta$ & Fraction effect within RA response window & 0.5\\ \hline 
$\xi$ & Fraction effect within CR timer & 0.5\\ \hline
\label{Summary of Parameters}
\end{tabular}
\end{table}

We evaluate the performance of the RA procedure within the time interval $T_{RAI}$, which ranges from the first preamble transmission of the first attempted UE to the completion of the RA procedure of the last attempted UE. $T_{RAI}$ is defined as \cite{9wei2015modeling}
\begin{equation}
T_{RAI}=T_{AP}+T_{W}+T_{RA},
\end{equation}
in which $T_{AP}$ denotes the arrival period, $T_{W}$ represents the average time used by the last UE to wait for the next available RA slot (unit: sub-frame) and $T_{RA}$ denotes the maximum period of time required by the last UE to complete the RA procedure. 

Represent $T_{RAI}$ in terms of the number of RA slots $K$, then $K$ is defined as $K = \left\lceil {\frac{{{T_{RAI}}}}{{{T_{RAP}}}}} \right\rceil$, where $T_{RAP}$ is the RA slot period.

\subsection{Preamble Transmission}
Let $R$ depict the total number of available preambles in a RA slot. Consider a specific preamble $r$ and let $Y_r^i$ be a random variable which takes value 1 if the preamble is used by exactly $i$ out of $m$ UEs and 0 otherwise. It readily follows that \cite{9wei2015modeling},
\begin{equation}
E\left[ {Y_r^i} \right] = \binom{m}{i}\frac{1}{{{R^i}}}{\left( {1 - \frac{1}{R}} \right)^{m - i}}.
\end{equation}
First, we consider the scenario where one preamble is used by two UEs, i.e. $i=2$.

The UEs are assumed to be uniformly distributed in the cell, thus the time interval between two UEs' arrivals $\Delta t$ is depicted as
\begin{equation}
\Delta t = \frac{{\left| {{d_1} - {d_2}} \right|}}{c},
\end{equation}
where $c = 3 \times {10^8}m/s$ while ${d_j}\left( {j = 1,2} \right)$ depicts the distance between the base station and the $j$-th UE, which follows the distribution given by \cite{chun2016stochastic}
\begin{equation}
{f_{d_j}}(x) = \frac{{2x}}{{{d_c}^2}},0 < x < {d_c},
\label{distance_distribution}
\end{equation}
where ${d_c}$ denotes the cell radius.

From Eq. \eqref{distance_distribution}, the distribution of $\Delta t$ can be readily computed and given by
\begin{equation}
{f_{\Delta t}}(y) = \frac{{4c}}{{3d_c^4}}\left( {2d_c^3 - 3d_c^2cy + {c^3}{y^3}} \right),{\rm{ }}0 < y < \frac{{{d_c}}}{c}.
\end{equation}

Then the probability of scenario 1 and 2 depicted in Fig. 3 can be expressed as follows,
\begin{equation}
\begin{aligned}
{p^{s1}} & = P\left\{ {\Delta t < {t_{rms}}} \right\} = \int_0^{{t_{rms}}} {{f_{\Delta t}}(y)dy} \\ & = \frac{{4c}}{{3d_c^4}}\left( {2d_c^3{t_{rms}} - \frac{3}{2}d_c^2ct_{rms}^2 + \frac{1}{4}{c^3}t_{rms}^4} \right)
\\
{p^{s2}} & = P\left\{ {\Delta t \ge {t_{rms}}} \right\} \\ & = 1 - \frac{{4c}}{{3d_c^4}}\left( {2d_c^3{t_{rms}} - \frac{3}{2}d_c^2ct_{rms}^2 + \frac{1}{4}{c^3}t_{rms}^4} \right)
\end{aligned},
\end{equation}
where $t_{rms}$ is the root meam square (RMS) of the delay spread. $p^{s2}$ represents the probability of differentiating the preamble transmissions of two UEs which select the same preamble. Let $p^{id}$ denote the probability of successfully separating the $i$-th UE' preamble signals within the ZCZ for $i \ge 3$. Based on the proposed NORA mechanism, the expected number of successful transmissions is given as
\begin{equation}
E\left[ S \right] = R\left( {E\left[ {Y_r^1} \right] + {p^{s2}}E\left[ {Y_r^2} \right]{\rm{ + }}{p^{3d}}E\left[ {Y_r^3} \right]{\rm{ + }}{p^{4d}}E\left[ {Y_r^4} \right] +  \cdots } \right).
\end{equation}
It is worth noting that for $i \ge 3$, the values of $E\left[ {Y_r^i} \right]$ are relatively small\footnote{The values of $E\left[ {Y_r^i} \right]$ are for $i \ge 4$ are less than 0.013. As for $E\left[ {Y_r^3} \right]$, the value can be comparably large when the number of attempt UEs is extremely high (the maximum value is 0.056 when the total number of arrived UEs is 50000), but it is still less than one third of $E\left[ {Y_r^2} \right]$.}. Furthermore, the derivation of $p^{id}$ is quite similar to the analysis process of $p^{s2}$. To illustrate the analytical model more briefly and coherently, we focus on the scenario of two UEs selecting the same preamble in subsequent discussions. Thus, $E\left[ S \right]$ is derived as
\begin{equation}
\begin{aligned}
E\left[ S \right] & = R\left( {E\left[ {Y_r^1} \right] + {p^{s2}}E\left[ {Y_r^2} \right]} \right) \\ & = m\left( {1 + \frac{{{p^{s2}}\left( {m - 1} \right)}}{{2\left( {R - 1} \right)}}} \right){\left( {1 - \frac{1}{R}} \right)^{m - 1}}
\end{aligned}.
\label{E[S]}
\end{equation}

\begin{figure*}[!t]
\centering
\includegraphics[width=0.8\linewidth]{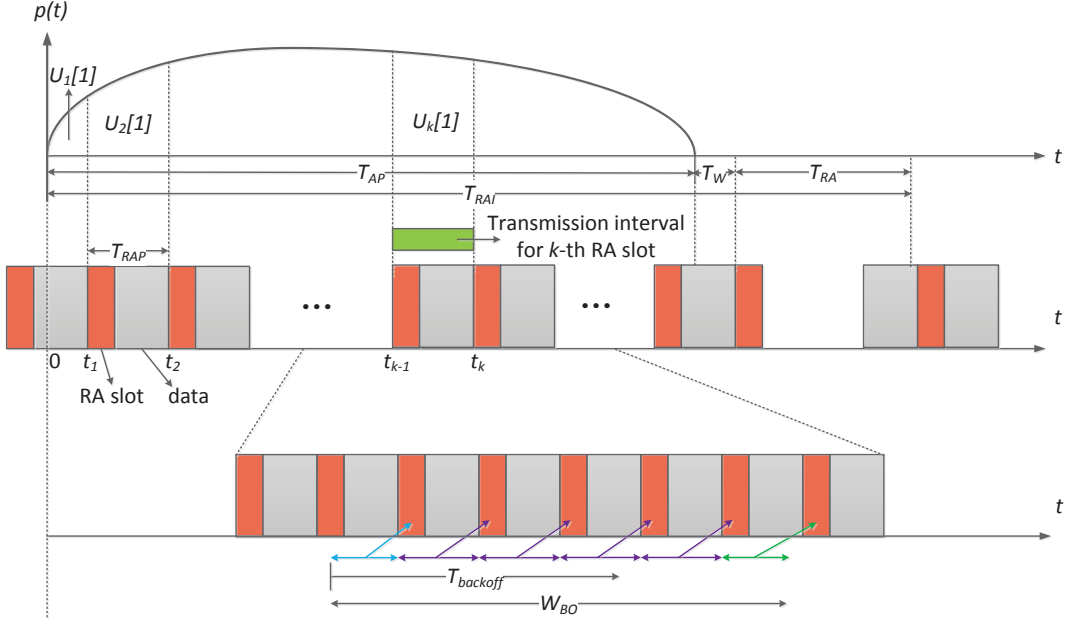}
\caption{The time diagram of the random access process.}
\label{Fig. 4}
\end{figure*}

Assume that ${U_k}[l]$ users attempt $l$-th $(l = 1, \ldots ,L)$ preamble transmission in the $k$-th $(k = 1, \ldots ,K)$ RA slot. As a result, a total of $U_{k}({U_k} = \sum\nolimits_{l = 1}^L {{U_k}\left[ l \right]})$ users transmit their preambles in the $k$-th RA slot. Thus, the expected number of UEs which successfully transmitted their preambles can be well approximated\footnote{For small $x$, ${e^{ - x}} = 1 - x$ (first term of the Taylor Expansion around point 0). Thus, for large $R$, ${e^{ - {U_k}/R}} = {\left( {{e^{ - 1/R}}} \right)^{{U_k}}} \approx {\left( {1 - 1/R} \right)^{{U_k}}} \approx {\left( {1 - 1/R} \right)^{{U_k} - 1}}$.} by 
\begin{equation}
E\left[ {S\left| {{U_k}} \right.} \right] \approx \left( {1 + \frac{{{p^{s2}}\left( {{U_k} - 1} \right)}}{{2\left( {R - 1} \right)}}} \right){U_k}{e^{ - \frac{{{U_k}}}{R}}}.
\label{UEswithPS}
\end{equation}

Taking the preamble detection probability $p_l$ of the $l$-th preamble transmission into account, the expected number of UEs whose $l$-th preamble transmissions are successfully detected in the $k$-th RA slot can be expressed as 
\begin{equation}
\begin{aligned}
{U_{k,PS}}\left[ l \right] & = \frac{{E\left[ {S\left| {{U_k}} \right.} \right]}}{{{U_k}}}{U_k}\left[ l \right]{p_l} \\ & \approx \left( {1 + \frac{{{p^{s2}}\left( {{U_k} - 1} \right)}}{{2\left( {R - 1} \right)}}} \right){U_k}\left[ l \right]{p_l}{e^{ - \frac{{{U_k}}}{R}}}
\end{aligned}.
\end{equation}

Thus, the total number of UEs whose preambles are detected in the $k$-th RA slot is given by 
\begin{equation}
\begin{aligned}
{U_{k,PS}} & = \sum\limits_{l = 1}^L {{U_{k,PS}}\left[ l \right]}  \\ & \approx \left( {1 + \frac{{{p^{s2}}\left( {{U_k} - 1} \right)}}{{2\left( {R - 1} \right)}}} \right)\left( {\sum\limits_{l = 1}^L {{U_k}\left[ l \right]{p_l}} } \right){e^{ - \frac{{{U_k}}}{R}}}
\end{aligned}.
\label{UEswithPS}
\end{equation}

Furthermore, the number of UEs which fail the $l$-th preamble transmissions is calculated as
\begin{equation}
\begin{aligned}
{U_{k,PF}}\left[ l \right] & = {U_k}\left[ l \right] - {U_{k,PS}}\left[ l \right] \\ & \approx \left( {1 - \left( {1 + \frac{{{p^{s2}}\left( {{U_k} - 1} \right)}}{{2\left( {R - 1} \right)}}} \right){p_l}{e^{ - \frac{{{U_k}}}{R}}}} \right){U_k}\left[ l \right]
\end{aligned}.
\end{equation}

\subsection{Message Transmission}
According to Section II. C, the UEs with successful preamble transmission will receive the RAR message and transmit the initial layer 3 message. In particular, the UEs in a NORA group will transmit their messages in the same resource blocks. However, due to channel distortion, the decoding of the layer 3 message may not be successful. As a result, the UEs with unsuccessful message transmission will return to preamble transmission. 

The UEs which have successfully transmitted their preambles are divided into two parts, i.e.
\begin{equation}
{U_{k,PS}}\left[ l \right] = U_{k,PS}^1\left[ l \right] + U_{k,PS}^2\left[ l \right],
\end{equation}
where $U_{k,PS}^1\left[ l \right]$ corresponds to UEs with no collision and $U_{k,PS}^2\left[ l \right]$ corresponds to UEs in NORA groups, which are readily calculated using Eg. \eqref{UEswithPS} as
\begin{align}
U_{k,PS}^1\left[ l \right] & = {U_k}\left[ l \right]{p_l}{e^{ - \frac{{{U_k}}}{R}}}. \\
U_{k,PS}^2\left[ l \right] & = \frac{{{p^{s2}}\left( {{U_k} - 1} \right)}}{{2\left( {R - 1} \right)}}{U_k}\left[ l \right]{p_l}{e^{ - \frac{{{U_k}}}{R}}}.
\end{align}

\setcounter{mytempeqncnt}{\value{equation}}
\setcounter{equation}{18}    
\begin{figure*}[!t]
\normalsize
\begin{equation}
\begin{split}
{U_{k,MS}}\left[ l \right] & = \left( {1 - {p_{out,0}}} \right)U_{k,PS}^1\left[ l \right] + \left( {1 - \frac{{{p_{out,1}} + {p_{out,2}}}}{2}} \right)U_{k,PS}^2\left[ l \right] \\
& = {e^{ - \frac{{{\phi _0}}}{{2{\theta ^2}}}}}{U_k}\left[ l \right]{p_l}{e^{ - \frac{{{U_k}}}{R}}} + \frac{{{\alpha _1}}}{2}{e^{ - \frac{{{\phi _1}}}{{2{\theta ^2}}}}}\left( {1 + {e^{ - \frac{{{\phi _2}}}{{2{\theta ^2}}}}}} \right)\frac{{{p^{s2}}\left( {{U_k} - 1} \right)}}{{2\left( {R - 1} \right)}}{U_k}\left[ l \right]{p_l}{e^{ - \frac{{{U_k}}}{R}}}.
\end{split}
\label{UMS}
\end{equation}
\hrulefill
\vspace*{4pt}
\end{figure*}
\setcounter{equation}{\value{mytempeqncnt}}

\setcounter{mytempeqncnt}{\value{equation}}
\setcounter{equation}{19}
\begin{figure*}[!t]
\normalsize
\begin{equation}
\begin{split}
{U_{k,MF}}\left[ l \right] & = {U_{k,PS}}\left[ l \right] - {U_{k,MS}}\left[ l \right] \\
& = \left( {1 - {e^{ - \frac{{{\phi _0}}}{{2{\theta ^2}}}}}} \right){U_k}\left[ l \right]{p_l}{e^{ - \frac{{{U_k}}}{R}}} + \left( {1 - \frac{{{\alpha _1}}}{2}{e^{ - \frac{{{\phi _1}}}{{2{\theta ^2}}}}}\left( {1 + {e^{ - \frac{{{\phi _2}}}{{2{\theta ^2}}}}}} \right)} \right)\frac{{{p^{s2}}\left( {{U_k} - 1} \right)}}{{2\left( {R - 1} \right)}}{U_k}\left[ l \right]{p_l}{e^{ - \frac{{{U_k}}}{R}}}.
\end{split}
\label{UMF}
\end{equation}
\hrulefill
\vspace*{4pt}
\end{figure*}
\setcounter{equation}{\value{mytempeqncnt}}

Then the number of UEs with successful message transmission is given in Eq. \eqref{UMS} while the number of UEs which succeed in preamble transmission but fail message transmission is readily shown in Eq. \eqref{UMF}. $p_{out,0}$ denotes the outage probability of the UEs with no collision while $p_{out,j}(j=1,2)$ depicts the outage probability of the $j$-th decoded UE in NORA groups (see Appendix for detailed derivation). 

\subsection{Random Backoff}
As illustrated in Fig. 4, the number of UEs which conduct their first preamble transmission in the $k$-th RA slot is given by
\setcounter{equation}{20}  
\begin{equation}
{U_k}[1] = U\int_{{t_{k - 1}} + 1}^{{t_k} + 1} {p(t)dt},
\end{equation}
where $p(t)$ is the arrival distribution and $t_k$ is the start of the $k$-th RA slot.
$p(t)$ follows that $\int_0^{{T_{AP}}} {p(t)dt}  = 1$.

The UEs with preamble or message transmission failure will perform random backoff before returning to preamble transmission. The number of contending UEs that transmit their $l$-th ($l \ge 2$) preamble in the $k$-th RA slot contains two parts. The first part originates from the UEs whose ($l-1$)-th preamble transmission failed (i.e. ${{U_{{k^{'}},PF}}[l - 1]}$) in the $k^{'}$ RA slot. Among these faild UEs, ${{p_{{k^{'}},k}}}$ of them end up transmitting the $l$-th preamble in the $k$-th RA slot after the random backoff process. Since these UEs perform uniform backoff within the backoff window $W_{BO}$ (length determined by BI in the RAR message), the value of ${{p_{{k^{'}},k}}}$ differs regarding ${k^{'}}(k_{\min }^{'} \le {k^{'}} \le k_{\max }^{'})$. The second part originates from the UEs that successfully transmit the ($l-1$)-th preamble in the $k$-th RA slot but fail the subsequent message transmission (i.e. ${U_{{k^{''}},MF}}[l - 1]$). ${{p_{{k^{''}},k}}}$ of the failed UEs will conduct the $l$-th preamble transmission in the $k^{''}$-th RA slot. ${{p_{{k^{''}},k}}}$ also has multiple probabilities regarding $k^{''} (k_{\min }^{''} \le {k^{''}} \le k_{\max }^{''})$.

Accordingly, for $2 \le l \le L$, ${U_k}[l]$ is expressed as
\begin{equation}
\begin{split}
{U_k}[l] = & \sum\limits_{{k^{'}} = k_{\min }^{'}}^{k_{\max }^{'}} {{p_{{k^{'}},k}}{U_{{k^{'}},PF}}[l - 1]} \\ 
& + \sum\limits_{{k^{''}} = k_{\min }^{''}}^{k_{\max }^{''}} {{p_{{k^{''}},k}}{U_{{k^{''}},MF}}[l - 1]}
\end{split},
\end{equation}
where ${U_{{k^{'}},PF}}[l - 1]$ represents the number of UEs which fail the ($l-1$)-th preamble transmission in the  ${k^{'}}$-th RA slot while ${U_{{k^{''}},MF}}[l - 1]$ denotes the number of UEs which succeed in the ($l-1$)-th preamble transmission in the ${k^{''}}$-th RA slot but fail the subsequent message transmission.

\begin{figure*}[!t]
\centering
\includegraphics[width=0.8\linewidth]{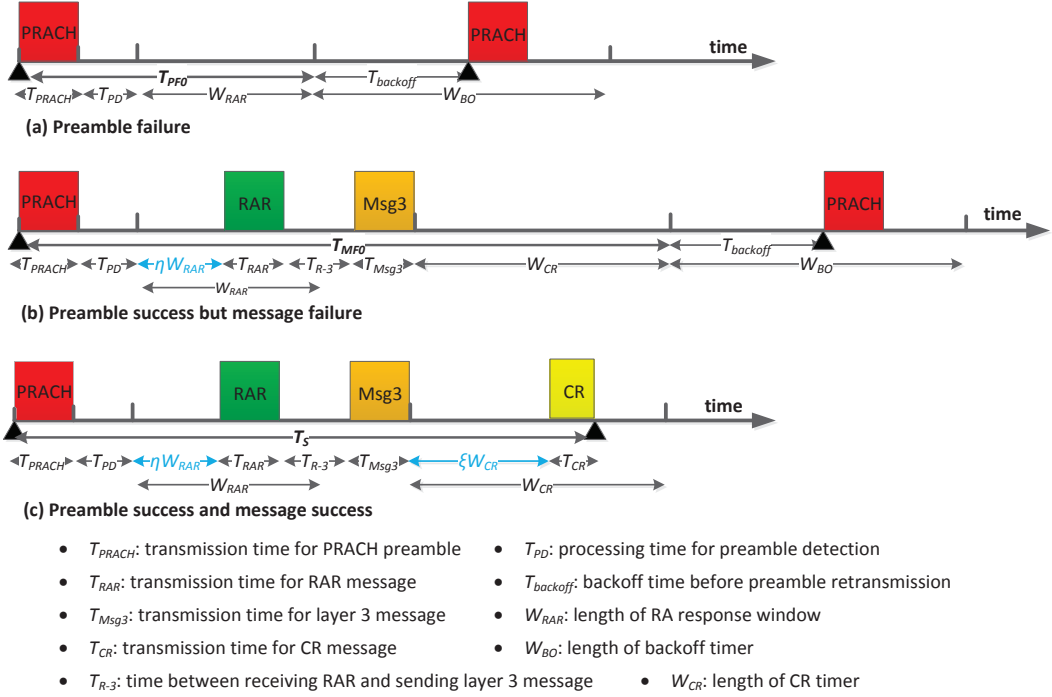}
\caption{The delay analysis regarding (a) preamble failure, (b) preamble success but message failure and (c) preamble success and message success.}
\label{Fig. 5}
\end{figure*}

As shown in Fig. \ref{Fig. 5}, the UEs which experience preamble failure in the ${k^{'}}$-th RA slot will realize their failure after ${T_{PF0}}$ subframes and perform random backoff. They will transmit the $l$-th preamble when the backoff timer expires. The backoff interval of the ${k^{'}}$-th RA slot starts from ${t_{{k^{'}}}} + {T_{PF0}} + 1$ and ends at ${t_{{k^{'}}}} + {T_{PF0}} + {W_{BO}}$. The UE will transmit their $l$-th preamble at the $k$-th RA slot if their backoff timer reach zero within the transmission interval of the $k$-th RA slot (from the start of the $(k-1)$-th RA slot to the start of the $k$-th RA slot, i.e. $\left[ {{t_{k - 1}},{t_k}} \right]$). $k_{\min }^{'}$ is obtained when the right-side boundary of the backoff interval of the $k^{'}$-th RA slot reaches the left-side boundary of the transmission interval of the $k$-th RA slot. Meanwhile, $k_{\max }^{'}$ is obtained when the left-side boundary of the backoff interval of the $k^{'}$-th RA slot reaches the right-side boundary of the transmission interval of the $k$-th RA slot. $k_{\min }^{'}$ and $k_{\max }^{'}$ are expressed as
\begin{equation}
k_{\min }^{'} = \left\lfloor {\left( {k - 1} \right) - \frac{{{T_{PF0}} + {W_{BO}}}}{{{T_{RAP}}}}} \right\rfloor.
\end{equation}
\begin{equation}
k_{\max }^{'} = \left\lceil {k - \frac{{{T_{PF0}}}}{{{T_{RAP}}}}} \right\rceil.
\end{equation}

${p_{{k^{'}},k}}$ is the percentage of the backoff interval of the preamble transmission in the $k^{'}$-th RA slot that overlaps with the transmission interval of the $k$-th RA slot. ${p_{{k^{'}},k}}$ is given by Eq. \eqref{overlap_percentage'}.

\setcounter{mytempeqncnt}{\value{equation}}
\setcounter{equation}{24}
\begin{figure*}[!t]
\normalsize
\begin{equation}
{p_{{k^{'}},k}} = \left\{ {\begin{array}{*{20}{c}}
{\begin{array}{*{20}{c}}
{\begin{array}{*{20}{c}}
{\frac{{\left( {{t_{{k^{'}}}} + {T_{PF0}} + {W_{BO}}} \right) - {t_{k - 1}}}}{{{W_{BO}}}},}\\
{\frac{{{T_{RAP}}}}{{{W_{BO}}}},}
\end{array}}\\
{\begin{array}{*{20}{c}}
{\frac{{{t_k} - \left( {{t_{{k^{'}}}} + {T_{PF0}}} \right)}}{{{W_{BO}}}},}\\
{0,}
\end{array}}
\end{array}}&{\begin{array}{*{20}{c}}
{\left( {k - 1} \right) - \frac{{{T_{PF0}} + {W_{BO}}}}{{{T_{RAP}}}} \le {k^{'}} \le k - \frac{{{T_{PF0}} + {W_{BO}}}}{{{T_{RAP}}}}}\\
{k - \frac{{{T_{PF0}} + {W_{BO}}}}{{{T_{RAP}}}} < {k^{'}} < \left( {k - 1} \right) - \frac{{{T_{PF0}}}}{{{T_{RAP}}}}}\\
{\left( {k - 1} \right) - \frac{{{T_{PF0}}}}{{{T_{RAP}}}} \le {k^{'}} \le k - \frac{{{T_{PF0}}}}{{{T_{RAP}}}}}\\
{{\rm{ortherwise}}}
\end{array}}
\end{array}} \right.
\label{overlap_percentage'}
\end{equation}
\hrulefill
\vspace*{4pt}
\end{figure*}
\setcounter{equation}{\value{mytempeqncnt}}

As illustrated in Fig. 5, the UEs which succeed in the ($l-1$)-th preamble transmission in the $k^{''}$-th RA slot but fail subsequent message transmission will perform random backoff when the CR timer expires. Thus, the backoff interval of the $k^{''}$-th RA slot starts from ${t_{{k^{''}}}} + {T_{MF0}} + 1$  and ends at ${t_{{k^{''}}}} + {T_{MF0}} + {W_{BO}}$. Similarly, $k_{\min }^{''}$ and $k_{\max }^{''}$ can be obtained as
\setcounter{equation}{25}  
\begin{equation}
k_{\min }^{''} = \left\lfloor {\left( {k - 1} \right) - \frac{{{T_{MF0}} + {W_{BO}}}}{{{T_{RAP}}}}} \right\rfloor.
\end{equation}
\begin{equation}
k_{\max }^{''} = \left\lceil {k - \frac{{{T_{MF0}}}}{{{T_{RAP}}}}} \right\rceil.
\end{equation}

The probability of the corresponding backoff interval falling into the transmission interval of the $k$-th RA slot is given by Eq. \eqref{overlap_percentage''}.
\setcounter{mytempeqncnt}{\value{equation}}
\setcounter{equation}{27}
\begin{figure*}[!t]
\normalsize
\begin{equation}
{p_{{k^{''}},k}} = \left\{ {\begin{array}{*{20}{c}}
{\begin{array}{*{20}{c}}
{\frac{{\left( {{t_{{k^{''}}}} + {T_{MF0}} + {W_{BO}}} \right) - {t_{k - 1}}}}{{{W_{BO}}}},}\\
{\frac{{{T_{RAP}}}}{{{W_{BO}}}},}\\
{\frac{{{t_k} - \left( {{t_{{k^{''}}}} + {T_{MF0}}} \right)}}{{{W_{BO}}}},}\\
{0,}
\end{array}}&{\begin{array}{*{20}{c}}
{\left( {k - 1} \right) - \frac{{{T_{MF0}} + {W_{BO}}}}{{{T_{RAP}}}} \le {k^{''}} \le k - \frac{{{T_{MF0}} + {W_{BO}}}}{{{T_{RAP}}}}}\\
{k - \frac{{{T_{MF0}} + {W_{BO}}}}{{{T_{RAP}}}} < {k^{''}} < \left( {k - 1} \right) - \frac{{{T_{MF0}}}}{{{T_{RAP}}}}}\\
{\left( {k - 1} \right) - \frac{{{T_{MF0}}}}{{{T_{RAP}}}} \le {k^{''}} \le k - \frac{{{T_{MF0}}}}{{{T_{RAP}}}}}\\
{{\rm{ortherwise}}}
\end{array}}
\end{array}} \right.
\label{overlap_percentage''}
\end{equation}
\hrulefill
\vspace*{4pt}
\end{figure*}
\setcounter{equation}{\value{mytempeqncnt}}

\subsection{Delay Analysis}
Define $\overline {{T_l}}$ as the average access delay of a successfully accessed UE that transmits exactly $l$ preambles. Based on the RA process proposed in Section II, $\overline {{T_l}}$ contains two parts. The first part originates from the time spent on $l-1$ failed preamble or message transmissions while the other parts originates from the time consumed by the $l$-th successful preamble and message transmission. Thus,
\setcounter{equation}{28}  
\begin{equation}
\overline {{T_l}}  = \left( {l - 1} \right)\overline {{T_F}}  + {T_S},
\label{T_l}
\end{equation}
where $\overline {{T_F}}$ denotes the average time of one failed preamble or message transmission while ${T_S}$ represents the time needed for one successful preamble and message transmission. 

Two scenarios exist in the event with one failed preamble or message transmission: preamble transmission fails or preamble transmission succeeds but message transmission fails. Therefore, $\overline {{T_F}}$ is computed as
\begin{equation}
\overline {{T_F}}  = \frac{{{p_{PF}}\overline {{T_{PF}}} }}{{{p_{PF}} + \left( {1 - {p_{PF}}} \right){p_{MF}}}} + \frac{{\left( {1 - {p_{PF}}} \right){p_{MF}}\overline {{T_{MF}}} }}{{{p_{PF}} + \left( {1 - {p_{PF}}} \right){p_{MF}}}},
\end{equation}
where ${p_{PF}}$ denotes the average probability of preamble transmission failure and ${p_{MF}}$ denotes the average probability of message transmission failure after successful preamble transmission. The derivation of ${p_{PF}}$ and ${p_{MF}}$ are given by
\begin{equation}
{p_{PF}} = {{\sum\limits_{k = 1}^K {\sum\limits_{l = 1}^L {{U_{k,PF}}\left[ l \right]} } } \mathord{\left/
 {\vphantom {{\sum\limits_{k = 1}^K {\sum\limits_{l = 1}^L {{U_{k,PF}}\left[ l \right]} } } {\sum\limits_{k = 1}^K {\sum\limits_{l = 1}^L {{U_k}\left[ l \right]} } }}} \right.
 \kern-\nulldelimiterspace} {\sum\limits_{k = 1}^K {\sum\limits_{l = 1}^L {{U_k}\left[ l \right]} } }}.
\end{equation}
\begin{equation}
{p_{MF}} = {{\sum\limits_{k = 1}^K {\sum\limits_{l = 1}^L {{U_{k,MF}}\left[ l \right]} } } \mathord{\left/
 {\vphantom {{\sum\limits_{k = 1}^K {\sum\limits_{l = 1}^L {{U_{k,MF}}\left[ l \right]} } } {\sum\limits_{k = 1}^K {\sum\limits_{l = 1}^L {{U_{k,PS}}\left[ l \right]} } }}} \right.
 \kern-\nulldelimiterspace} {\sum\limits_{k = 1}^K {\sum\limits_{l = 1}^L {{U_{k,PS}}\left[ l \right]} } }}.
\end{equation}

$\overline {{T_{PF}}}$ and $\overline {{T_{MF}}}$ represents the average time required for a failed preamble transmission and a failed message transmission with successful preamble transmission respectively, as depicted in Eq. \eqref{T_PF} and Eq. \eqref{T_MF}. As illustrated in Fig. \ref{Fig. 5}, $\overline {{T_{PF}}}$ contains the transmission time for PRACH preamble, processing time for the base station to detect preamble, waiting time before the RA response window ends and the backoff time before preamble retransmission. Regarding $\overline {{T_{MF}}}$, with the successful preamble transmission, the UE is able to receive the RAR message before the ${W_{RAR}}$ timer ends. $\eta (0 < \eta  < 1)$ is introduced to model this effect. Compared to $\overline {{T_{PF}}}$, $\overline {{T_{MF}}}$ adds the transmission time for RAR and layer 3 message, the time interval between a UE receiving a RAR and sending layer 3 message as well as waiting time before the CR timer expires. 
\begin{equation}
\overline {{T_{PF}}}  = {T_{PF0}} + \overline {{T_{backoff}}},
\label{T_PF}
\end{equation}
\begin{equation}
\overline {{T_{MF}}}  = {T_{MF0}} + \overline {{T_{backoff}}},
\label{T_MF}
\end{equation}
where 
\begin{equation}
{T_{PF0}} = {T_{PRACH}} + {T_{PD}} + {W_{RAR}}.
\end{equation}
\begin{equation}
\begin{split}
{T_{MF0}} = & {T_{PRACH}} + {T_{PD}} + \eta {W_{RAR}} + {T_{RAR}} \\ 
& + {T_{R - 3}} + {T_{Msg3}} + {W_{CR}}
\end{split}.
\end{equation}

As for ${T_S}$, since the UE is capable of receiving the CR message before the CR timer expires, the fraction effect $\xi (0 < \xi  < 1)$ is considered. ${T_S}$ contains the transmission time for CR message, as shown in Eq. \eqref{T_S}.
\begin{equation}
\begin{split}
{T_S} = & {T_{PRACH}} + {T_{PD}} + \eta {W_{RAR}} + {T_{RAR}} \\ 
& + {T_{R - 3}} + {T_{Msg3}} + \xi {W_{CR}} + {T_{CR}}
\end{split}.
\label{T_S}
\end{equation}

Since UEs perform random uniform backoff, ${T_{backoff}}$ follows a uniform distribution within $\left[ {0,{W_{BO}}} \right]$. Thus,
\begin{equation}
\overline {{T_{backoff}}}  = \frac{1}{2}{W_{BO}}.
\end{equation}

\section{Performance Evaluation}
\subsection{Performance Metrics}
\subsubsection{Throughput of Preamble Transmission}
Consider a certain RA slot in which a total number of $m$ UEs transmit their preambles. The throughput of preamble transmission $R_P$ is defined as the number of successful transmitted preambles given $m$ simultaneous preamble transmissions. In the NORA scheme, it's readily given from Eq. \eqref{E[S]} as
\begin{equation}
R_P^{NORA} = E\left[ S \right]{\rm{ = }}m\left( {1 + \frac{{{p^{s2}}\left( {m - 1} \right)}}{{2\left( {R - 1} \right)}}} \right){\left( {1 - \frac{1}{R}} \right)^{m - 1}}.
\end{equation}

In the ORA scheme, the throughput is expressed as
\begin{equation}
R_P^{ORA} = m{\left( {1 - \frac{1}{R}} \right)^{m - 1}}
\end{equation}
which states that a UE succeeds in the preamble transmission if all the other $m-1$ UEs select the other $R-1$ preambles.

As illustrated in Fig. 6, the maximal throughput of preamble transmission 20 is achieved when $m$ equals to 53 or 54 in the ORA scheme. In comparison, more than 30\% of throughput improvement can be achieved when $m$ equals to 69 in the NORA scheme. 
\begin{figure}[!t]
\centering
\includegraphics[width=0.7\linewidth]{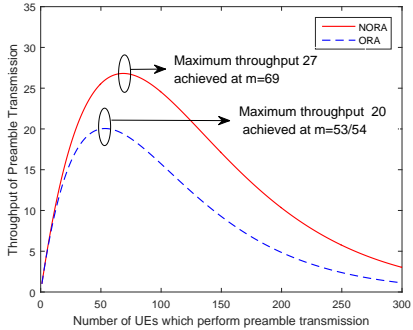}
\caption{Throughput of preamble transmission given ${p^{s2}} = 0.6$ and $R = 54$.}
\label{Fig. 6}
\end{figure}

\subsubsection{Throughput of RA Process}
${R_{RA}}$ is defined as the total number of successfully accessed UEs in the RA time interval $T_{RAI}$, i.e. maximum supported UEs. That is, ${R_{RA}}$ is the accumulated number of UEs with successful preamble and message transmission, expressed as
\begin{equation}
{R_{RA}} = \sum\limits_{k = 1}^K {\sum\limits_{l = 1}^L {{U_{k,MS}}\left[ l \right]} }.
\end{equation}

\subsubsection{Collision Probability}
${P_C}$ is defined as the ratio between the number of occurrences when two and more UEs make a RA attempt using exactly the same preamble (resulting in undetected collision) and the overall number RA preambles (with or without access attempts) in the period. That is, ${P_C}$ is the ratio between the number of undetected collided preambles and the overall number of preambles. The number of undetected collided preambles is equal to the total number of preambles minus the number of idle preambles and successfully received  preambles. Hence, ${P_C}$ is given as
\begin{equation}
{P_C} = {{\sum\limits_{k = 1}^K {\left( {R - R\left( {E\left[ {Y_r^0\left| {{U_k}} \right.} \right]} \right) - {U_{k,PS}}} \right)} } \mathord{\left/
 {\vphantom {{\sum\limits_{k = 1}^K {\left( {R - R\left( {E\left[ {Y_r^0\left| {{U_k}} \right.} \right]} \right) - {U_{k,PS}}} \right)} } {KR}}} \right.
 \kern-\nulldelimiterspace} {KR}}.
\end{equation}

\subsubsection{Access Success Probability}
${P_S}$ is defined as the probability that a UE successfully completes the RA procedure within the maximum number of preamble transmissions. That is, ${P_S}$ is the ratio between total number of successfully accessed UEs and the total number of UEs arrived in ${T_P}$, expressed as
\begin{equation}
{P_S} = {{{R_{RA}}} \mathord{\left/
 {\vphantom {{{R_{RA}}} U}} \right.
 \kern-\nulldelimiterspace} U}.
\end{equation}

\subsubsection{CDF of the Number of Preamble Transmissions}
$F(m)$ is defined as the CDF of the number of preamble transmissions to perform a RA procedure for the successfully accessed UEs, where $m$ is the number of preamble transmissions. More specifically, it is the ratio between the number of successfully accessed UEs which transmit no more than $m$ preambles and the number of all successfully accessed UEs, given as
\begin{equation}
F(m) = {{\sum\limits_{k = 1}^K {\sum\limits_{l = 1}^m {{U_{k,MS}}\left[ l \right]} } } \mathord{\left/
 {\vphantom {{\sum\limits_{k = 1}^K {\sum\limits_{l = 1}^m {{U_{k,MS}}\left[ l \right]} } } {\sum\limits_{k = 1}^K {\sum\limits_{l = 1}^L {{U_{k,MS}}\left[ l \right]} } }}} \right.
 \kern-\nulldelimiterspace} {\sum\limits_{k = 1}^K {\sum\limits_{l = 1}^L {{U_{k,MS}}\left[ l \right]} } }}.
\end{equation}

\subsubsection{CDF of the Access Delay}
$G(d)$ is defined as the CDF of the delay between the first preamble attempt and the completion of the RA process for the successfully accessed UEs. That is, $G(d)$ is the ratio between the number of the successfully accessed UEs whose access delay is no greater than $d$ and the total number of successfully accessed UEs. $G(d)$ is estimated by 
\begin{equation}
G(d) = {{\sum\limits_{k = 1}^K {\sum\limits_{l = 1}^{{m_{\max }}(d)} {{U_{k,MS}}\left[ l \right]} } } \mathord{\left/
 {\vphantom {{\sum\limits_{k = 1}^K {\sum\limits_{l = 1}^{{m_{\max }}(d)} {{U_{k,MS}}\left[ l \right]} } } {\sum\limits_{k = 1}^K {\sum\limits_{l = 1}^L {{U_{k,MS}}\left[ l \right]} } }}} \right.
 \kern-\nulldelimiterspace} {\sum\limits_{k = 1}^K {\sum\limits_{l = 1}^L {{U_{k,MS}}\left[ l \right]} } }},
\end{equation}
where ${m_{\max }}(d)({m_{\max }}(d) \in N)$ is the maximal number of preamble transmissions by an UE and is estimated by setting $l = {m_{\max }}(d)$ in Eq. \eqref{T_l} and let $\overline {{T_l}}=d$.

\subsubsection{Average Number of Preamble Transmissions for Successfully Accessed UEs}
$\overline L$ is defined as the ratio between the total number of preamble transmissions for all the successfully accessed UEs and the total number of all the successfully accessed UEs. Hence, $\overline L$ is given as
\begin{equation}
\bar L = {{\sum\limits_{k = 1}^K {\sum\limits_{l = 1}^m {{U_{k,MS}}\left[ l \right] \cdot l} } } \mathord{\left/
 {\vphantom {{\sum\limits_{k = 1}^K {\sum\limits_{l = 1}^m {{U_{k,MS}}\left[ l \right] \cdot l} } } {\sum\limits_{k = 1}^K {\sum\limits_{l = 1}^L {{U_{k,MS}}\left[ l \right]} } }}} \right.
 \kern-\nulldelimiterspace} {\sum\limits_{k = 1}^K {\sum\limits_{l = 1}^L {{U_{k,MS}}\left[ l \right]} } }}.
\end{equation}

\subsubsection{Average Access Delay for Successfully Accessed UEs}
$\overline {D_{RA}}$ is defined as the ratio between the total access delay of preamble transmissions for all the successfully accessed UEs and the total number of all the successfully accessed UEs. $D_{RA}$ is estimated by 
\begin{equation}
\overline {{D_{RA}}}  = {{\sum\limits_{k = 1}^K {\sum\limits_{l = 1}^{{m_{\max }}(d)} {{U_{k,MS}}\left[ l \right]}  \cdot \overline {{T_l}} } } \mathord{\left/
 {\vphantom {{\sum\limits_{k = 1}^K {\sum\limits_{l = 1}^{{m_{\max }}(d)} {{U_{k,MS}}\left[ l \right]}  \cdot \overline {{T_l}} } } {\sum\limits_{k = 1}^K {\sum\limits_{l = 1}^L {{U_{k,MS}}\left[ l \right]} } }}} \right.
 \kern-\nulldelimiterspace} {\sum\limits_{k = 1}^K {\sum\limits_{l = 1}^L {{U_{k,MS}}\left[ l \right]} } }}.
\end{equation}

Among the aforementioned performance metrics, the throughput of RA process ${R_{RA}}$, collision probability ${P_C}$, access success probability ${P_S}$ and average access delay $\overline {{D_{RA}}}$ are of paramount importance to RA performance analysis and scheme design. ${R_{RA}}$ serves as the indicator of maximum network load, while ${P_C}$ and ${P_S}$ measures the reliability of network access. $\overline {{D_{RA}}}$ provides the benchmark for network access delay, which is critical for delay-sensitive services.

\subsection{Simulation Analysis}
\begin{figure}[!t]
\centering
\includegraphics[width=0.7\linewidth]{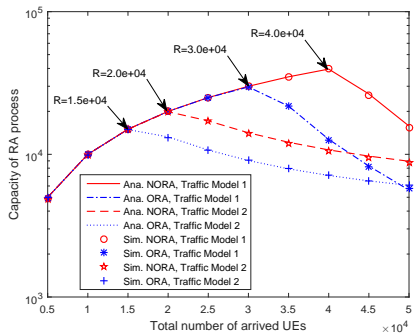}
\caption{Throughput Of RA process of the NORA and ORA schemes under both Traffic Models.}
\label{Fig. 7}
\end{figure}

Simulations are performed on a C-based platform to validate the correctness of the analytical model. In the following figures, lines denote the analytical results while symbols depict the simulation results. Each point in the simulation results represents the average value of $10^4$ samples of the RA process outcome. The simulation parameter configurations for NORA performance evaluation is specified in bold form in Table I. 
The uniform and beta preamble arrival distributions \cite{6TR37.868} are considered within the period [0, $T_{AP}$]. Regarding the uniform distribution, $p(t) = {{{T_{RAP}}} \mathord{\left/
 {\vphantom {{{T_{RAP}}} {{T_{AP}}}}} \right.
 \kern-\nulldelimiterspace} {{T_{AP}}}}$.
As for the beta distribution,
\begin{equation}
p(t) = \frac{{{t^{\alpha  - 1}}{{({T_{AP}} - t)}^{\beta  - 1}}}}{{T_{AP}^{^{\alpha  + \beta  - 1}}Beta(\alpha ,\beta )}},\alpha  = 3,\beta  = 4,
\end{equation}
where $Beta(\alpha ,\beta )$ is the beta function.

As demonstrated in Fig. 7, the throughput of the NORA and ORA schemes under Traffic Model 1 are 40000 and 30000, respectively. However, the corresponding values are reduced to 20000 and 15000 under Traffic Model 2 for that it is considered as an extreme scenario in which a large amount of UEs access the network in a highly synchronized manner. Nevertheless, the throughput of RA process in the NORA scheme exhibits a 30\% advantage compared to ORA under both Traffic Models.

\begin{figure} 
\centering 
\subfigure[Traffic Model 1]{
\label{fig:subfig:a} 
\includegraphics[width=0.7\linewidth]{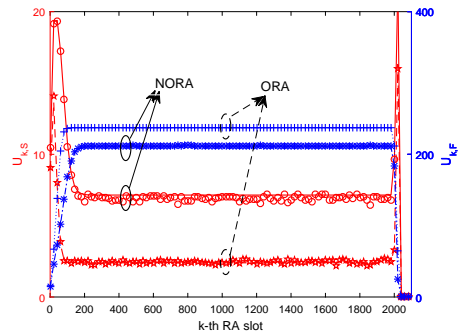}} \hspace{0.01\linewidth} 
\subfigure[Traffic Model 2]{
\label{fig:subfig:b} 
\includegraphics[width=0.7\linewidth]{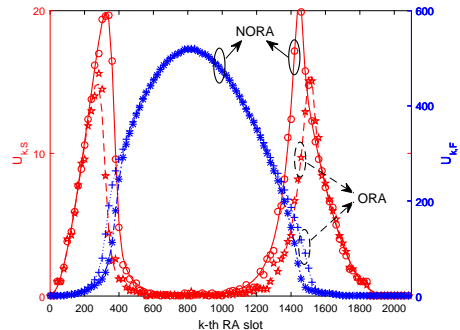}}
\caption{The number of succeeded and failed UEs in the $k$-th RA slot of the NORA and ORA schemes under both Traffic Models.} 
\label{PerSlot} 
\end{figure}

Fig. 8 shows the number of succeeded UEs (${U_{k,MS}} = \sum\nolimits_{l = 1}^L {{U_{k,MS}}\left[ l \right]}$) and failed UEs ($U_{k}-U_{k,MS}$) in the $k$-th RA slot of the NORA and ORA schemes under Traffic Model 1 and 2. $U=50000$ is taken to model the overloaded scenario. As shown in Fig. 8(a), the number of succeeded UEs of ORA and NORA both sharply increase and reach a peak at the beginning of the RA period. Then ORA precedes NORA in the swift fall of the number of succeeded UEs, which later remains steady except for a momentary spike at the end of the RA period. It is evident that the number of succeeded UEs in the NORA scheme is nearly three times of that in the ORA scheme during the stable phase, which results from the non-orthogonal characteristic feature in the preamble and message transmission. Moreover, ORA experienced an earlier saturation and larger number of failed UEs compared to NORA.  

Regarding Traffic Model 2, which is depicted in Fig. 8(b), the number of succeeded UEs for ORA scheme first demonstrates a constant growth thanks to random backoff algorithm and reaches a maximum value at $k=250$. In the meantime, the number of succeeded UEs for NORA scheme continues to rise until $k=300$. But then they are both significantly reduced to zero when $k$ increases from 500 to 1100 due to the excessive collisions resulted from the accumulated failed UEs. Nevertheless, they start to increase again after $k=1100$ since more UEs declare access failure after reaching the maximum number of preamble transmissions and stop contending for RA slots. The NORA scheme exhibits an obviously rapider growing trend until it reaches another maximum value at $k=15000$ and then decreases since fewer UEs attempt random access. As for the number of failed UEs, NORA and ORA go through the same stable phase but the value of ORA outside the stable phase is higher than NORA, which reflects that ORA undergoes more sever collision. 

\begin{figure} 
\centering 
\subfigure[Collision probability]{
\label{fig:subfig:a} 
\includegraphics[width=0.7\linewidth]{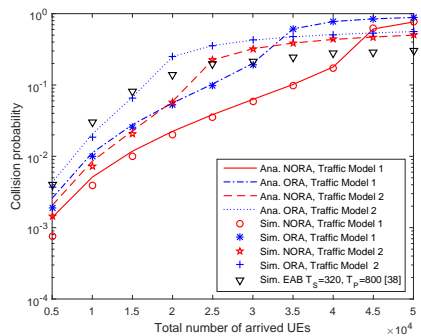}} \hspace{0.01\linewidth} 
\subfigure[Access success probability]{
\label{fig:subfig:b} 
\includegraphics[width=0.7\linewidth]{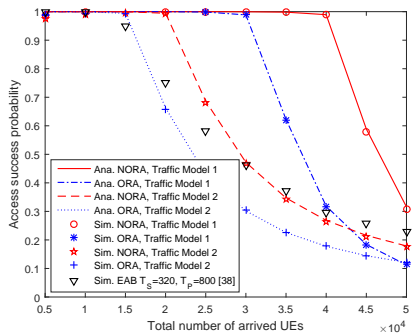}}
\caption{Collision and access success probability of the NORA and ORA schemes under both Traffic Models.} 
\label{CP&AP} 
\end{figure}

Fig. 9(a) compares the collision probability of the NORA and ORA schemes under both Traffic Models, while Fig. 9(b) depicts the access success probability. It is illustrated that the performance of RA process under Traffic Model 1 is better than Traffic Model 2, which coincides with the fact that Traffic Model 2 is an extreme scenario in which a large amount of UEs attempt to access the network in a highly synchronized manner. The collision probability shows a continuously growing trend with the increase of UE amount under both Traffic Models. Specifically, when the total number of UEs exceeds the maximum supported UEs (readily seen in Fig. 7), for instance 20000 for the NORA scheme and 15000 for the ORA scheme under Traffic Model 2, the growth rate of the probability witnesses a significant increase. Moreover, the difference of collision probability between the NORA and ORA schemes peaks in case of the total number of UEs being 35000 under Traffic Model 1 and 20000 under Traffic Model 2. Regarding Traffic Model 2, the NORA scheme can support up to 20000 UEs, which improves the number of maximum supported UEs by more than 30\% compared to the ORA scheme. It is worthy to note that when the number of arrived UEs exceeds the amount the network can endure, the access success probability sees a continuously sharp decline until it approaches zero. 
Performance comparisons with the state-of-the art RA scheme EAB \cite{Cheng2015Modeling} are also exhibited. Regarding collision and access probability, NORA far outperforms EAB when the total number of arrived UEs is smaller than 25000. The performance of EAB is superior to NORA when U exceeds 25000. Nevertheless, this performance benefit comes with the great expense of access delay, which will be further elaborated later.

\begin{figure} 
\centering 
\subfigure[Preamble transmission number]{
\label{fig:subfig:a} 
\includegraphics[width=0.7\linewidth]{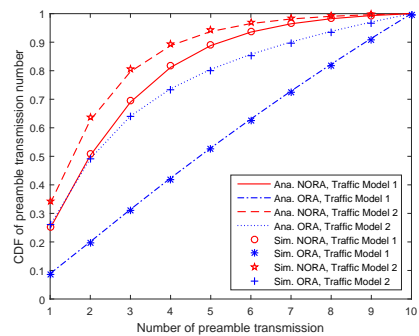}} \hspace{0.01\linewidth} 
\subfigure[Access delay]{
\label{fig:subfig:b} 
\includegraphics[width=0.7\linewidth]{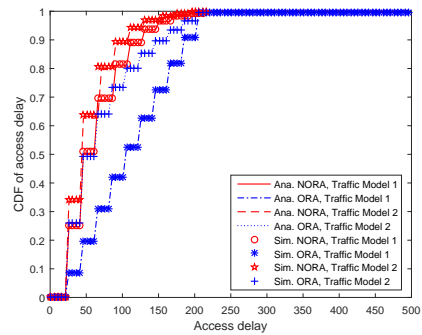}}
\caption{CDF of the number of preamble transmissions and access delay for the successfully accessed UEs in NORA and ORA schemes under both Traffic Models.} 
\label{CDF} 
\end{figure}

Fig. 10 shows the CDF of the number of preamble transmissions and the access delay for the successfully accessed UEs in both schemes under Traffic Model 1 and 2. The total number of arrived UEs is set as 40000 under Traffic Model 1 and 20000 under Traffic Model 2, which corresponds to the maximum supported UEs depicted in Fig. 7. Regarding the number of preamble transmissions, more than 25\% of the UEs can complete their random access with one-shot preamble transmission in NORA scheme while only 10\% can successfully access the network by transmitting one preamble under Traffic Model 1. Under Traffic Model 2, the corresponding figures are 33\% for the NORA scheme and 25\% for the ORA scheme, respectively. It can be seen that the reduction of the access delay is distinct under Traffic Model 1 while not particularly obvious under Traffic Model 2. This is for the reason that Traffic Model 1 characterizes a typical scenario in which UEs access the network in a non-synchronized manner while Traffic Model 2 is considered as a much more challenging scenario in which a large number of UEs attempt network access in a highly synchronized manner e.g. after a power outage. In future 5G networks, the extreme scenarios featured by Traffic Model 2 tend to be more dominant. Therefore, a more delicate NORA scheme is required against such challenging scenarios, which is also one of our main future works.

\begin{figure} 
\centering 
\subfigure[Average L]{
\label{fig:subfig:a} 
\includegraphics[width=0.7\linewidth]{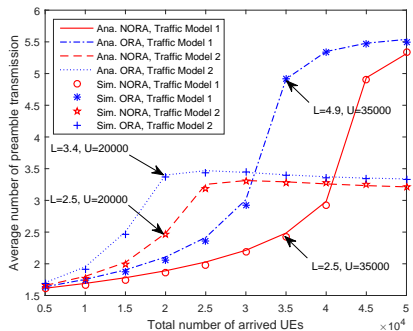}} \hspace{0.01\linewidth} 
\subfigure[Average access delay]{
\label{fig:subfig:b} 
\includegraphics[width=0.7\linewidth]{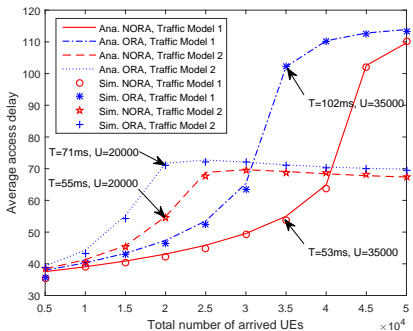}}
\caption{Average number of preamble transmissions and access delay for the successfully accessed UEs in NORA and ORA schemes under both Traffic Models.} 
\label{CDF} 
\end{figure}

\begin{table*}[!tp]  
\caption{Access delay comparison}
\newcommand{\tabincell}[2]{\begin{tabular}{@{}#1@{}}#2\end{tabular}}
\centering          
\begin{tabular}{ccccccccccc}
\hline
\diagbox[width=4em,trim=l]{$\overline {D_{RA}}$(ms)}{U} & 5000 & 10000 & 15000 & 20000 & 25000 & 30000 & 35000 & 40000 & 45000 & 50000 \\\hline
NORA & 38 & 40 & 45 & 55 & 68 & 69 & 69 & 69 & 68 & 67 \\\hline
ORA & 39 & 43 & 54 & 71 & 71 & 71 & 70 & 69 & 68 & 69 \\\hline
EAB & 31 & 3985 & 4654 & 4359 & 3949 & 3604 & 3023 & 2502 & 2166 & 1908 \\\hline
\end{tabular}
\end{table*}

The average number of preamble transmissions of NORA and ORA scheme is depicted in Fig. 11(a) regarding various load conditions (total number of arrived UEs). Under Traffic Model 1, the required number of preamble transmissions constantly rises with increasing load. NORA provides a significant reduction of preamble transmission times when $U>20000$. In particular, it manages to halve the required number of preamble transmissions when $U=35000$ compared to ORA. Under traffic model 2, NORA effectively cuts down the number of the preamble transmissions when the total number of arrived UEs $U$ is smaller than 25000 but provides limited advantage over ORA given more UEs. In addition, the average number of preamble transmissions for NORA and ORA both maintain at a relatively constant value when $U$ exceeds 25000. It results from the fact that the network is heavily congested for both NORA and ORA with more than 25000 UEs following beta distribution. Resembling conclusions can be obtained regarding average access delay from Fig. 11(b).

Moreover, the average access delay of NORA under Traffic Model 2 is compared with EAB, as demonstrated in Table II. It's reflected that the access delay of EAB stays at a very high level when U exceeds 5000. Specifically, the access delay of EAB is more than a hundred times larger than that of NORA when the total number of arrived UEs is 15000. Nevertheless, it is noticeable that the access delay of EAB gradually declines as the total number of UEs increases, which gives it superiority in extremely overloaded scenarios. For that reason, the integration of NORA and EAB against explosive access scenarios is one of our main research directions.

\section{Conclusion}
In this paper, we have proposed the NORA scheme to alleviate the potential access congestion problem regarding the massive-connection scenarios in 5G networks. Specifically, the spatial distribution characteristics of UEs were utilized to realize multi-preamble detection and RAR reception, which effectively improves the preamble transmission success probability. Moreover, NORA allows simultaneous message transmission of multiple UEs, thus alleviates the demand on limited PUSCH resources.
In addition, we have presented the analytical model to investigate the transient behavior of the NORA process with non-stationary arrivals under realistic assumptions. Besides, a comprehensive evaluation of our proposition is given, including throughput, access success probability, number of preamble transmission and access delay. Simulation results indicate that NORA outperforms ORA in terms of all the considered metrics, especially for a relatively large number of UEs (e.g. 50000 UEs). Compared with ORA, NORA can increase the throughput of the RA process by more than 30\%. Moreover, NORA manages to halve the required preamble transmissions and access delay when the total number of UEs is near the RA throughput.


%
\appendix[Derivation of $p_{out,1}$, $p_{out,2}$ and $p_{out,0}$]
Based on the conclusions in Section III. A, we assume that there are 2 UEs in a NORA group which transmit their layer 3 messages in the same resource blocks. Since the UEs perform timing alignment in accordance with the received TA value before message transmission, the received signal at the base station can be expressed as 
\begin{equation}
y = \sum\limits_{j = 1}^2 {{h_j}\sqrt {{P_j}} {x_j}}  + n,
\end{equation}
where $h_j$ denotes the channel between the $j(j=1,2)$-th UE and the base station, $P_j$ represents the transmit power of the $j$-th UE in a NORA group (see Eq. \eqref{transmit_power_i}) while $x_j$ denotes the layer 3 message for the $j$-th UE, and $n$ is the additive noise at the base station. 

Regarding $h_j$, the large scale path loss and Rayleigh fading are assumed. Therefore $h_i$ is expressed as ${h_j} = {l_j}{v_j}$. $l_j$ is expressed as ${l_j} = {{4\pi {d_j}} \mathord{\left/
 {\vphantom {{4\pi {d_j}} {\sqrt {{A_j}} \lambda }}} \right.
 \kern-\nulldelimiterspace} {\sqrt {{A_j}} \lambda }}$, where $d_j$ denotes the distance between the $j$-th UE and the base station, $A_j$ represents the product of the transmit and receive antenna losses and $\lambda$ denotes the wavelength.
$v_j$ denotes the Rayleigh channel gain with the probability density function (PDF) of ${\left| {{v_i}} \right|}$ given by
\begin{equation}
{f_{\left| {{v_j}} \right|}}(x) = \frac{x}{{{\theta ^2}}}{e^{ - \frac{{{x^2}}}{{2{\theta ^2}}}}},0 \le x < \infty,
\end{equation}
where $\theta ^2$ represents the variance. 

At the base station, the messages of UEs within a NORA group are decoded in a successive way based on the decoding order, i.e. power back-off order. Thus, the achievable data rate of the two UEs in a NORA group is given by Eq. \eqref{eq:rate1} and Eq. \eqref{eq:rate2} respectively.
\begin{equation}
{R_1} = \log \left( {1 + \frac{{{P_1}{{\left| {{h_1}} \right|}^2}}}{{{P_2}{{\left| {{h_2}} \right|}^2} + {\sigma ^2}}}} \right),
\label{eq:rate1}
\end{equation}
\begin{equation}
{R_2} = \log \left( {1 + \frac{{{P_2}{{\left| {{h_2}} \right|}^2}}}{{{\sigma ^2}}}} \right),
\label{eq:rate2}
\end{equation}
where $\sigma ^2$ denotes the variance of noise.

Let ${\hat R_j}$ denote the target data rate of the $j$-th UE. The successful detection of the $j$-th UE's message can be defined as ${Z_j}=\left\{{{R_j} \ge \hat R_j}\right\}$. Given that the detection of the $j$-th UE's message is based on the successful decoding of the prior $(j-1)$ UEs' messages, the outage probability of the $j$-th UE is expressed as \cite{zhang2016uplink}
\begin{equation}
{p_{out,j}} = 1 - P\left( {Z_1 \cap \cdots \cap Z_j} \right),
\end{equation}

Following the similar derivations in \cite{zhang2016uplink}, the outage probabilities of the 1-st and 2-nd decoded UEs in a NORA group are given as 
\begin{align}
{p_{out,1}} & = 1 - {\alpha _1}{e^{ - \frac{{{\phi _1}}}{{2{\theta ^2}}}}}, \\
{p_{out,2}} & = 1 - {\alpha _1}{e^{ - \frac{{{\phi _1} + {\phi _2}}}{{2{\theta ^2}}}}},
\end{align}
where ${\alpha _1} = \frac{2}{{1 + {{10}^{ - \frac{\delta }{{10}}}}\left( {{2^{{{\hat R}_1}}} - 1} \right)}}$ and ${\phi _j} = \frac{{{\sigma ^2}}}{{{P_j}l_j^2}}\left( {{2^{{{\hat R}_j}}} - 1} \right)$ for $j=1,2$.

Regarding the UEs with no collision, the outage probability is given as 
\begin{equation}
{p_{out,0}} = P\left\{ {\log \left( {1 + \frac{{{P_0}{{\left| {{h_0}} \right|}^2}}}{{{\sigma ^2}}}} \right) \ge {{\hat R}_0}} \right\} = 1 - {e^{ - \frac{{{\phi _0}}}{{2{\theta ^2}}}}},
\end{equation}
where ${\phi _0} = \frac{{{\sigma ^2}}}{{{P_0}l_0^2}}\left( {{2^{{{\hat R}_0}}} - 1} \right)$.

%

\small
\bibliographystyle{IEEEtran}   
\bibliography{references}

\begin{thebibliography}{10}
\providecommand{\url}[1]{#1}
\csname url@samestyle\endcsname
\providecommand{\newblock}{\relax}
\providecommand{\bibinfo}[2]{#2}
\providecommand{\BIBentrySTDinterwordspacing}{\spaceskip=0pt\relax}
\providecommand{\BIBentryALTinterwordstretchfactor}{4}
\providecommand{\BIBentryALTinterwordspacing}{\spaceskip=\fontdimen2\font plus
\BIBentryALTinterwordstretchfactor\fontdimen3\font minus
  \fontdimen4\font\relax}
\providecommand{\BIBforeignlanguage}[2]{{%
\expandafter\ifx\csname l@#1\endcsname\relax
\typeout{** WARNING: IEEEtran.bst: No hyphenation pattern has been}%
\typeout{** loaded for the language `#1'. Using the pattern for}%
\typeout{** the default language instead.}%
\else
\language=\csname l@#1\endcsname
\fi
#2}}
\providecommand{\BIBdecl}{\relax}
\BIBdecl

\bibitem{17NORA}
Y.~Liang, X.~Li, J.~Zhang, and Y.~Liu, ``{A novel random access scheme based on
  successive interference cancellation for 5G networks},'' in \emph{2017 IEEE
  Wireless Communications and Networking Conference (WCNC)}, Mar. 2017, pp.
  1--6.

\bibitem{16andrews2014will}
J.~G. Andrews, S.~Buzzi, W.~Choi, S.~V. Hanly, A.~Lozano, A.~C. Soong, and
  J.~C. Zhang, ``{What will 5G be?}'' \emph{IEEE J. Sel. Areas Commun.},
  vol.~32, no.~6, pp. 1065--1082, Jun. 2014.

\bibitem{12TS36.300}
\emph{Evolved Universal Terrestrial Radio Access (E-UTRA); Overall Description;
  Stage 2}, 3GPP TS 36.300 V13.2.0, Jan. 2016.

\bibitem{9wei2015modeling}
C.~H. Wei, G.~Bianchi, and R.~G. Cheng, ``{Modeling and analysis of random
  access channels with bursty arrivals in OFDMA wireless networks},''
  \emph{IEEE Trans. Wireless Commun.}, vol.~14, no.~4, pp. 1940--1953, Apr.
  2015.

\bibitem{1TS36.321}
\emph{Evolved Universal Terrestrial Radio Access (E-UTRA); Medium Access
  Control (MAC)}, 3GPP TS 36.321 V9.3.0, Jun. 2010.

\bibitem{2TS36.211}
\emph{Evolved Universal Terrestrial Radio Access (E-UTRA); Physical Channels
  and Modulation}, 3GPP TS 36.211 V10.4.0, Dec. 2011.

\bibitem{7LTE}
S.~Stefania, T.~Issam, and B.~Matthew, \emph{LTE-The UMTS Long Term Evolution:
  From Theory to Practice}.\hskip 1em plus 0.5em minus 0.4em\relax Wiley, 2011.

\bibitem{11laya2014random}
A.~Laya, L.~Alonso, and J.~Alonso~Zarate, ``{Is the random access channel of
  LTE and LTE-A suitable for M2M communications? A survey of alternatives},''
  \emph{IEEE Commun. Surveys Tuts.}, vol.~16, no.~1, pp. 4--16, Oct. 2014.

\bibitem{13wang2015optimal}
Z.~Wang and V.~W. Wong, ``{Optimal access class barring for stationary machine
  type communication devices with timing advance information},'' \emph{IEEE
  Trans. Wireless Commun.}, vol.~14, no.~10, pp. 5374--5387, Oct. 2015.

\bibitem{wiriaatmadja2015hybrid}
D.~T. Wiriaatmadja and K.~W. Choi, ``Hybrid random access and data transmission
  protocol for machine-to-machine communications in cellular networks,''
  \emph{IEEE Trans. Wireless Commun.}, vol.~14, no.~1, pp. 33--46, Jan. 2015.

\bibitem{duand}
S.~Duan, V.~Shah-Mansouri, Z.~Wang, and V.~Wong, ``{D-ACB: Adaptive congestion
  control algorithm for bursty M2M traffic in LTE networks},'' \emph{IEEE
  Trans. Veh. Technol}, vol.~65, no.~12, pp. 9847–--9861, Dec. 2016.

\bibitem{morvari2016two}
F.~Morvari and A.~Ghasemi, ``{Two-Stage resource allocation for random access
  M2M communications in LTE network},'' \emph{IEEE Commun. Lett.}, vol.~20,
  no.~5, pp. 982--985, May 2016.

\bibitem{seo2011design}
J.-B. Seo and V.~C. Leung, ``{Design and analysis of backoff algorithms for
  random access channels in UMTS-LTE and IEEE 802.16 systems},'' \emph{IEEE
  Trans. Veh. Technol}, vol.~60, no.~8, pp. 3975--3989, Oct. 2011.

\bibitem{lin2014prada}
T.-M. Lin, C.-H. Lee, J.-P. Cheng, and W.-T. Chen, ``{PRADA: Prioritized random
  access with dynamic access barring for MTC in 3GPP LTE-A networks},''
  \emph{IEEE Trans. Veh. Technol}, vol.~63, no.~5, pp. 2467--2472, Jun. 2014.

\bibitem{14pang2014network}
Y.~C. Pang, S.~L. Chao, G.~Y. Lin, and H.~Y. Wei, ``{Network access for M2M/H2H
  hybrid systems: A game theoretic approach},'' \emph{IEEE Commun. Lett.},
  vol.~18, no.~5, pp. 845--848, Jun. 2014.

\bibitem{15chen2015delayed}
J.~Chen, Y.~T. Lin, and R.~G. Cheng, ``{A delayed random access speed-up scheme
  for group paging in machine-type communications},'' in \emph{2015 IEEE
  International Conference on Communications (ICC)}, Jun. 2015, pp. 623--627.

\bibitem{6TR37.868}
\emph{Study on RAN Improvements for Machine-type Communications}, 3GPP TR37.868
  V11.0.0, Sep. 2011.

\bibitem{5Thomsen14code}
H.~Thomsen, N.~K. Pratas, {\v{C}}.~Stefanovi{\'c}, and P.~Popovski, ``{Code
  expanded radio access protocol for machine-to-machine communications},''
  \emph{Trans. Emerging Tel. Tech.}, vol.~24, no.~4, pp. 355–--365, Jun.
  2013.

\bibitem{4ko14spatial}
H.~S. Jang, S.~M. Kim, K.~S. Ko, J.~Cha, and D.~K. Sung, ``{Spatial group based
  random access for M2M communications},'' \emph{IEEE Commun. Lett.}, vol.~18,
  no.~6, pp. 961–--964, Jun. 2014.

\bibitem{3ko2012novel}
K.~S. Ko, M.~J. Kim, K.~Y. Bae, D.~K. Sung, J.~H. Kim, and J.~Y. Ahn, ``{A
  novel random access for fixed-location machine-to-machine communications in
  OFDMA based systems},'' \emph{IEEE Commun. Lett.}, vol.~16, no.~9, pp.
  1428--1431, Sep. 2012.

\bibitem{niyato2014performance}
D.~Niyato, P.~Wang, and D.~I. Kim, ``Performance modeling and analysis of
  heterogeneous machine type communications,'' \emph{IEEE Trans. Wireless
  Commun.}, vol.~13, no.~5, pp. 2836--2849, May 2014.

\bibitem{10kim2015enhanced}
T.~Kim, H.~S. Jang, and D.~K. Sung, ``{An enhanced random access scheme with
  spatial group based reusable preamble allocation in cellular M2M networks},''
  \emph{IEEE Commun. Lett.}, vol.~19, no.~10, pp. 1714--1717, Oct. 2015.

\bibitem{Yu2008High}
Y.~Yu and G.~B. Giannakis, ``High-throughput random access using successive
  interference cancellation in a tree algorithm,'' \emph{IEEE Trans. Inf.
  Theory}, vol.~53, no.~12, pp. 4628--4639, Dec. 2007.

\bibitem{Salvo2015Power}
P.~Salvo~Rossi, K.~Kansanen, R.~R. Muller, and C.~Rachinger, ``Power
  randomization for iterative detection over random-access fading channels,''
  \emph{IEEE Trans. Wireless Commun.}, vol.~14, no.~10, pp. 5704--5713, Jun.
  2015.

\bibitem{Applebaum2012Asynchronous}
L.~Applebaum, W.~U. Bajwa, M.~F. Duarte, and R.~Calderbank, ``Asynchronous
  code-division random access using convex optimization,'' \emph{Phys.
  Commun.}, vol.~5, no.~2, pp. 129--147, Jun. 2012.

\bibitem{Wang2008A}
X.~Wang, Y.~Yu, and G.~B. Giannakis, ``A robust high-throughput tree algorithm
  using successive interference cancellation,'' \emph{IEEE Trans. Commun.},
  vol.~55, no.~12, pp. 2253--2256, Dec. 2007.

\bibitem{Wang2008Design}
------, ``Design and analysis of cross-layer tree algorithms for wireless
  random access,'' \emph{IEEE Trans. Wireless Commun.}, vol.~7, no.~3, pp.
  909--919, Mar. 2008.

\bibitem{Zanella2012Theoretical}
A.~Zanella and M.~Zorzi, ``Theoretical analysis of the capture probability in
  wireless systems with multiple packet reception capabilities,'' \emph{IEEE
  Trans. Commun.}, vol.~60, no.~4, pp. 1058--1071, Feb. 2012.

\bibitem{Lin2015A}
H.~Lin, K.~Ishibashi, W.~Shin, and T.~Fujii, ``A simple random access scheme
  with multilevel power allocation,'' \emph{IEEE Commun. Lett.}, vol.~19,
  no.~12, pp. 2118--2121, Oct. 2015.

\bibitem{Xu2013Decentralized}
C.~Xu, L.~Ping, P.~Wang, S.~C. Chan, and X.~Lin, ``Decentralized power control
  for random access with successive interference cancellation,'' \emph{IEEE J.
  Sel. Areas Commun.}, vol.~31, no.~11, pp. 2387--2396, Nov. 2013.

\bibitem{ding2014performance}
Z.~Ding, Z.~Yang, P.~Fan, and H.~V. Poor, ``{On the performance of
  non-orthogonal multiple access in 5G systems with randomly deployed users},''
  \emph{IEEE Signal Process. Lett.}, vol.~21, no.~12, pp. 1501--1505, Dec.
  2014.

\bibitem{ding2015cooperative}
Z.~Ding, M.~Peng, and H.~V. Poor, ``{Cooperative non-orthogonal multiple access
  in 5G systems},'' \emph{IEEE Commun. Lett.}, vol.~19, no.~8, pp. 1462--1465,
  Jun. 2015.

\bibitem{zhang2016uplink}
N.~Zhang, J.~Wang, G.~Kang, and Y.~Liu, ``{Uplink nonorthogonal multiple access
  in 5G systems},'' \emph{IEEE Commun. Lett.}, vol.~20, no.~3, pp. 458--461,
  Mar. 2016.

\bibitem{liu2016cooperative}
Y.~Liu, Z.~Ding, M.~Elkashlan, and H.~V. Poor, ``Cooperative non-orthogonal
  multiple access with simultaneous wireless information and power transfer,''
  \emph{IEEE J. Sel. Areas Commun.}, vol.~34, no.~4, pp. 938--953, Apr. 2016.

\bibitem{1TS36.331}
\emph{Evolved Universal Terrestrial Radio Access (E-UTRA); Radio Resource
  Control (RRC)}, 3GPP TS 36.331 V12.4.1, Dec. 2014.

\bibitem{6TR36.213}
\emph{Evolved Universal Terrestrial Radio Access (E-UTRA); Physical Layer
  Procedures}, 3GPP TR36.213 V12.4.0, Dec. 2014.

\bibitem{chun2016stochastic}
Y.~J. Chun, S.~L. Cotton, H.~S. Dhillon, A.~Ghrayeb, and M.~O. Hasna, ``A
  stochastic geometric analysis of device-to-device communications operating
  over generalized fading channels,'' \emph{arXiv:1605.03244}, May 2016.

\bibitem{Cheng2015Modeling}
R.~G. Cheng, J.~Chen, D.~W. Chen, and C.~H. Wei, ``{Modeling and analysis of an
  extended access barring algorithm for machine-type communications in LTE-A
  networks},'' \emph{IEEE Trans. Wireless Commun.}, vol.~14, no.~6, pp.
  2956--2968, Jun. 2015.

\end{thebibliography}

\end{document}